# Bicritical and tetracritical phenomena and scaling properties of the SO(5) theory

Xiao Hu

*National Research Institute for Metals, Tsukuba 305-0047, Japan*

By large scale Monte Carlo simulations it is shown that the stable fixed point of the SO(5) theory is either bicritical or tetracritical depending on the effective interaction between the antiferromagnetism and superconductivity orders. There are no fluctuation-induced first-order transitions suggested by $\epsilon$ expansions. Bicritical and tetracritical scaling functions are derived for the first time and critical exponents are evaluated with high accuracy. Suggestions on experiments are given.

High-$T_c$ superconductivity (SC) in cuprates is achieved by carrier doping into the half-filling insulator when the antiferromagnetism (AF) is suppressed. The SO(5) symmetry between AF and $d$-wave SC has been proposed by Zhang which should be achieved at the bicritical point where the two critical (*second-order*) lines merge [1]. The variation of the chemical potential from the bicritical point breaks the SO(5) symmetry into the SO(3) for AF and the U(1) for SC. The question on the asymmetry between AF and SC caused by the Hubbard gap has been resolved by introducing the projected SO(5) concept [2]. Reasons that the SO(5) symmetry can be realized in reality have been discussed successfully by Arrigoni and Hanke [3]. Although the SO(5) theory gives a natural explanation on the phase diagrams observed experimentally, there is still no direct observation of the bicritical point in cuprates, partially because of the difficulty of chemical preparation of samples. Observations favoring the SO(5) scenario are reported in other systems: a bicritical point in a set of organic superconductors [4], and more recently a tetracritical point in heavy-fermion superconductors [5], mainly from the shape of phase diagrams. Since experimental data are piling in a wide range of materials, it becomes important to test the SO(5) theory in real systems in an unambiguous way, such as measuring the critical exponents [6,7]. To make this approach feasible in practice, and to resolve more fundamental problems discussed below, better theoretical understanding about the bicritical phenomenon of the SO(5) theory is necessary.

Theoretical studies on bicritical phenomena go back to more than two decades ago [8]. According to RG $\epsilon$ expansions a bicritical point becomes instable when the degrees of freedom $n$ exceeds a critical value, $n_c \simeq 4$ in three dimensions (3D), and a so-called tetracritical point with four critical lines crossing takes over [9]. Since $n = 5$ in the SO(5) theory is only slightly above $n_c$, the stable fixed point is nontrivial. As a matter of fact, basing on $\epsilon$ expansions Murakami and Nagaosa argued [6] that the bicritical point occurs only exceptionally; instead one should observe either a tetracritical point or a triple point where two *first-order* lines merge into a single first-order line (see Ref. [10] for a more general discussion about possible phase diagrams). Since the SO(5) symmetry is achieved only at the bicritical point, a full understand on its stability is of fundamental importance for an assessment of the SO(5) theory.

The RG is not necessarily correct in 3D when $\epsilon$ expansions are truncated at low orders. A particularly relevant example is the normal to SC transition, which is fluctuation-driven first order in $d = 4 - \epsilon$, but is in XY-like second order in 3D. In order to derive the correct fixed points of the SO(5) theory by a strong-coupling approach, we have performed large-scale, classical Monte Carlo (MC) simulations. The main results are as follows: (1) The bicritical (tetracritical) point is stable for repulsive (attractive) interactions between the AF and SC orders. There are no fluctuation-induced first-order transitions suggested by $\epsilon$ expansions; (2) Bicritical and tetracritical scaling functions are derived for the first time; (3) The critical exponents and the ratios between coefficients of the two critical lines are evaluated with high accuracy. Since these properties are universal for the SO(5) theory, microscopic details are irrelevant. This opens the way to test unambiguously the SO(5) theory by experiments.

The Hamiltonian in the present study is [11]

$$\mathcal{H} = -J \sum_{\langle i,j \rangle} \mathbf{s}_i \cdot \mathbf{s}_j - J \sum_{\langle i,j \rangle} \mathbf{t}_i \cdot \mathbf{t}_j + g \sum_i \mathbf{s}_i^2 + w \sum_i \mathbf{s}_i^2 \mathbf{t}_i^2, \tag{1}$$

on the simple cubic lattice [12]. The coupling is limited to nearest neighbors. The vector $\mathbf{s}(\mathbf{t})$ of three(two) components is for the local AF(SC) order parameter, and $\mathbf{s}_i^2 + \mathbf{t}_i^2 = 1$ [13]. The symmetry-breaking $g$ field is related to energy gaps of magnons and charged pairs, and the chemical potential of electron by $g = (\Delta_s - \Delta_c + \mu)/2$ [2]. The $w$ field governs the interaction between the AF and SC orders. Since terms of $\sum_i \mathbf{s}_i^4$ and $\sum_i \mathbf{t}_i^4$ can be mapped into the $g$ and $w$ terms up to a constant, and higher order terms are irrelevant to the critical phenomena, the above Hamiltonian is general enough for investigation on the multicritical phenomena of the SO(5) theory [14]. In the above Hamiltonian, $w >, =, < 0$ corresponds to $w^2 >, =, < uv$ in Ref. [6], respectively.

In a typical simulation process, we generate a random configuration of superspins at a sufficiently high temperature, and then cool down the system gradually using the standard Metropolis algorithm of the MC technique. The $g$ and $w$ fields are fixed in this simulation process. Typically, we use 50,000 MC steps for equilibration and 100,000 MC steps for statistics on each site in the system at each temperature. The system size is $L^3 = 40^3$ with periodic boundary conditions. We have also simulated systems of $L^3 = 60^3$ for several parameter sets and confirmed that the finite-size effects do not change the results.

In order to establish the scaling analysis and to make accurate estimates of critical exponents, let us start with $w = 0$, taking advantage of the exact bicritical value $g_b = 0$. The temperature dependence of the helicity modulus [15,16], which is proportional to the superfluid density and thus serves as the long-range order parameter of SC, is shown in Fig. 1(a) for positive $g$ fields. For negative $g$ fields we cannot observe the SC long-range order. The critical temperature and the critical exponent defined in $\Upsilon \sim (1 - T/T_c)^\upsilon$ are evaluated for $g = 0.5J$ as $\upsilon = 0.666 \pm 0.004$ and $T_c = 0.9773 \pm 0.0004 J/k_B$ as shown in Fig. 1(a), where data points in $0.8J/k_B \leq T \leq 0.96J/k_B$ are used [17]. This estimate of the critical exponent is in good agreement with $\nu_2 = 0.669 \pm 0.002$ by RG for the *pure* 3D XY (n=2) model [18] noting $\upsilon = (D-2)\nu_2$ where $D = 3$. Therefore, at least for $g \geq 0.5J$ there is a sufficiently large critical region where AF fluctuations are irrelevant to the SC critical phenomenon.

The temperature dependence of the Néel order parameter is shown in Fig. 1(b) for negative $g$ fields. We cannot observe the AF long-range order for any positive $g$ field. The Néel point and the critical exponent defined in $m_3 \equiv \frac{1}{N}\sqrt{< \sum_i \mathbf{s}_i^2 >} \sim (1 - T/T_N)^{\beta_3}$ are evaluated as $\beta_3 = 0.363 \pm 0.001$ and $T_N = 0.9357 \pm 0.0002 J/k_B$ for $g = -0.5J$ as shown in Fig. 1(b), where data points in temperature region $0.7J/k_B \leq T \leq 0.92J/k_B$ are used. This estimate is in good agreement with the value $\beta_3 = 0.3645 \pm 0.0025$ by RG for the *pure* 3D Heisenberg (n=3) model [18]. Therefore, at least for $g \leq -0.5J$ there is a sufficiently large critical region where SC fluctuations are irrelevant to the AF critical phenomenon.

As seen in Fig. 1, the critical onsets of the order parameters below the corresponding critical points become steeper in both SC and AF regimes as the absolute value of the $g$ field is reduced. In order to analyze the crossover phenomena accurately, we extend the scaling theory for divergent quantities *above* the critical lines [19] to order parameters *below* the critical lines. We hypothesize that the helicity modulus for $T < T_c(g)$ where $g > g_b = 0$ should depend on temperature and the $g$ field in a scaled way

$$\Upsilon(T; g) \simeq A(g - g_b)^{\nu_s/\phi} \times f[(T/T_b - 1)/(g - g_b)^{1/\phi}] \qquad (2)$$

with $\nu_5$ the critical exponent for correlation length for $n = 5$ and the crossover exponent $\phi$. The scaling function $f(x)$ should have the following properties: (i) $f(x) \simeq (-x)^{\nu_5}$, as $x \to -\infty$ (i.e. $g \to g_b$ for $T < T_b$), which covers the critical phenomenon at $g = g_b$; (ii) $f(x) \simeq (B_2 - x)^{\nu_2}$, as $x \to B_2$ with $B_2$ defined in $B_2 \times (g - g_b)^{1/\phi} = T_c(g)/T_b - 1$ (i.e. $T \to T_c(g)$), which accommodates the singularities at $T_c(g)$'s for $g > g_b$. From the scaling ansatz (2), one should have

$$\Upsilon(T_b; g)/\Upsilon(T_b; g') = [(g - g_b)/(g' - b_b)]^{\nu_5/\phi}. \tag{3}$$

Therefore, the bicritical point can be estimated as the temperature where the ratio on the left-hand side depends merely on the ratio $(g - g_b)/(g' - b_b)$ but not on $g$ fields themselves. Meanwhile, the combination of exponents $\nu_5/\phi$ can be determined. From the temperature dependence of the ratio $\Upsilon(T; g)/\Upsilon(T; g')$ displayed in Fig. 2(a) for $[g_1, g_1'] = [0.05J, 0.10J]$ and $[g_2, g_2'] = [0.10J, 0.20J]$ with $g_1/g_1' = g_2/g_2' = 0.5$ ($g_b = 0$), we estimate $T_b \simeq 0.8458 \pm 0.0005J/k_B$ and $\nu_5/\phi \simeq 0.525 \pm 0.002$. Each point in Fig. 2(a) is obtained by 3 million MC steps on each superspin in the system.

The crossover exponent $\phi$ can be evaluated from the slopes $\kappa_1$ of $\Upsilon(T; g_1)/\Upsilon(T; g_1')$ and $\kappa_2$ of $\Upsilon(T; g_2)/\Upsilon(T; g_2')$ with respect to temperature at the bicritical point $T_b$:

$$\phi = \ln[(g_2 - g_b)/(g_1 - g_b)]/\ln(\kappa_1/\kappa_2). \tag{4}$$

Fitting the temperature dependence of $\Upsilon(T; g)/\Upsilon(T; g')$ up to the second order of deviation from the bicritical temperature as shown by the solid curves in Fig. 2(a), one obtains $\kappa_1 \simeq 7.7301 \pm 0.0804$ and $\kappa_2 \simeq 4.6907 \pm 0.0309$. Therefore, we have $\phi = 1.387 \pm 0.030$ with the error bar estimated from those of the slopes. Finally, we arrive at $\nu_5 = 0.728 \pm 0.018$.

Our estimates of the critical exponents should be compared with the results by the $\epsilon$ expansions of RG [20,21] $\nu_5 = 0.723$ (up to the third order of $\epsilon$) and $\phi = 1.313$ (up to the second order of $\epsilon$). The agreement between the two estimates on $\nu_5$ is extremely fine. The present estimate on $\phi$ is expected to be of higher precision than that of the $\epsilon$ expansions.

The scaling plot of Eq.(2) for all data in Fig. 1(a) is displayed in the main panel of Fig. 2. A magnification around zero helicity modulus is given in Fig. 2(b), where the solid curve corresponds to the critical exponent $\nu_2 = 0.666$ and $B_2 \simeq 1/4$. It is clear that both properties (i) and (ii) are satisfied very well.

It is worthy to point out that, using the scaling property discussed above, the bicritical point and the associated bicritical exponents can be inferred from informations away from the bicritical point.

The same scaling analysis can be performed for the Néel order parameter for $g < 0$ in Fig. 1(b). Since the Néel order parameter suffers more serious finite-size effects, the plot similar to Fig. 2(a) fails to give accurate estimate of critical exponents. Alternatively, we evaluate the order parameter $m_5 \equiv \frac{1}{N}\sqrt{< \sum_i \mathbf{s}_i^2 > + < \sum_i \mathbf{t}_i^2 >}$ at $g = g_b = 0$ by MC simulation as shown in Fig. 3(a). The critical point and critical exponent are evaluated as $\beta_5 = 0.400 \pm 0.002$ and $T_b = 0.8559 \pm 0.0003J/k_B$, using data points in temperature region $0.7J/k_B \leq T \leq 0.84J/k_B$. The value of the critical exponent agrees very well with $\beta = 0.402$ by the $\epsilon$ expansions to second order [20]. A fine scaling plot is displayed in the main panel of Fig. 3. From Fig. 3(b), one reads $B_3 \simeq 1/6$ with $B_3$ defined in $B_3 \times (g_b - g)^{1/\phi} = T_N(g)/T_b - 1$. The ratio $B_2/B_3 = 3/2$ is given very accurately by the inverse ratio between degrees of freedom (discussed in Ref. [10] in a different context) and should be a universal constant for the SO(5) theory.

We then turn to $w > 0$ for which the fixed point is nontrivial as discussed previously. Without losing generality, let us set $w = 0.1J$. We observe the SC long-range order for $g \geq 0.012J$ and AF one for $g \leq 0.010J$ as shown in Figs. 4(a) and 5(a). At $g = 0.011J$ either the AF or the SC order is obtained depending on the initial random configuration and cooling process, same as at $g = g_b = 0$ for $w = 0$. As clearly seen in Figs. 4(a) and 5(a), all the phase transitions upon temperature reduction are second order, and thus there are no fluctuation-induced first-order transitions suggested by $\epsilon$ expansions. No coexistence between AF and SC is observed in the $g - T$ plane, and the SC and AF long-range order parameters reach their full values at zero temperature as in Figs. 4(a) and 5(a) (see Figs. 6(a) and (b) for the case of coexistence). Therefore, the fixed point for $w = 0.1J$ is a bicritical point at $g_b = 0.011J$ and $T_b = 0.8458J/k_B$ (equal to that for $w = 0$ because of the same exchange coupling). Furthermore, all the data in Figs. 4(a) and 5(a) collapse on the bicritical scaling curves in the main panels of Figs. 4 and 5 with the same bicritical exponents obtained for $w = 0$, except for some deviations observed in the Néel order parameter for $g = 0.010J$ coming from the finite-size effect. We observe the universal ratio $B_2/B_3 = 3/2$ in the present case too. From these observations, we conclude that the phase diagram for $w \geq 0$ is governed by a bicritical point.

For the negative cubic anisotropy $w = -0.1J$, we observe a tetracritical point at $g_t = -0.011J$ (of the same absolute value of $g_b$ for $w = 0.1J$) around which AF and SC coexist in a narrow region in the phase diagram displayed in Fig. 6. From the scaling analysis shown in Fig. 7(a) with $[g_1, g_1'] = [0.039J, 0.089J]$ and $[g_2, g_2'] = [0.089J, 0.189J]$ with $(g_1 - g_t)/(g_1' - g_t) = (g_2 - g_t)/(g_2' - g_t) = 0.5$, we obtain $T_t = 0.8460 \pm 0.0005J/k_B$ (same with $T_b$ for $w \geq 0$ within the error bar), $\nu_t/\phi_t \simeq 0.529 \pm 0.002$, and $\kappa_1 \simeq 7.8900 \pm 0.0935$ and $\kappa_2 \simeq 4.7451 \pm 0.0250$. Adopting the same process developed for $w = 0$, we arrive at $\nu_t \simeq 0.721 \pm 0.018$, $\phi_t \simeq 1.363 \pm 0.030$, and $\beta_t \simeq 0.399 \pm 0.001$, where $\beta_t$ is evaluated from data in Fig. 8(a). The critical and crossover exponents for the bicritical and tetracritical phenomena are too close to each other to be differed numerically within the errors of estimation. It should be difficult for experiments too. Our estimates on the tetracritical critical and crossover exponent are more accurate than the $\epsilon$-expansion results $\nu_t = 0.632$ and $\phi_t = 1.155$ to the first order of $\epsilon$ [6,9]. Deviations from the scaling curves in the main panels of Figs. 7 and 8 occur when the AF and SC orders coexist. The universal ratio $B_2/B_3 = 3/2$ is observed once again.

$\rho - T$ phase diagram with the charge density $\rho = \langle \mathbf{t}_i^2 \rangle/2$ can also be composed such as the one shown in Fig. 1(c) for $w = 0$. The two lines below the bicritical point are given by $\rho_{\pm 0}(T) = \lim_{g \to \pm 0} \rho(T; g)$. The phase surrounded by these two curves corresponds to $g = g_b = 0$. In this phase SC and AF coexist in a macroscopic level, while in a mesoscopic or microscopic level there can be phase separations, such as stripes. The discontinuous jump in the charge density at $g = g_b = 0$ and $T < T_b$ is described numerically by $\rho_{+0}(T) - \rho_{-0}(T) \sim (1 - T/T_b)^{\beta_\rho}$. We obtain $\beta_\rho = 0.80 \pm 0.08$ using the scaling relations $\beta_\rho = 2 - \alpha_5 - \phi$ [8] and $\alpha_5 = 2 - D\nu_5 = -0.184 \pm 0.054$. As temperature is reduced at $g = g_b = 0$, the charge compressibility diverges as $\partial \rho/\partial g \sim (T - T_b)^{-\gamma_\rho}$ with $\gamma_\rho = 2\phi - 2 + \alpha_5 \simeq 0.59 \pm 0.11$. This divergent charge compressibility might be responsible for the peculiar transport properties observed below the pseudogap temperature in cuprates. The shapes of the critical lines in the $\rho - T$ phase diagram Fig. 1(c) are also consistent with the behaviors $\pm(\rho - 0.2) \sim (T_{c,N}/T_b - 1)^{\beta_\rho}$ as predicted by the scaling theory [8].

Finally, a suggestion for experimental investigations is made. As revealed in our study, by means of the scaling properties it is able to infer the bicritical (or tetracritical) point and the associated critical phenomena even in cases that sample preparations are difficult in the merging region of the AF and SC orders, such as high-$T_c$ cuprates. Therefore, systematic measurements on the London penetration depth (which is related to the SC order parameter discussed in the present study by $\lambda \sim 1/\sqrt{\Upsilon}$) for a series of dopings near the boundary to the AF phase (see Ref. [22] for optimal doping) should be able to clarify the symmetry governing the high-$T_c$ superconductivity.

---

Figure Captions

Fig. 1: $g - T$ phase diagram with a bicritical point for $w = 0$. Inset (a): temperature dependence of the helicity moduli; inset (b): that of the Néel order parameters; inset (c): $\rho - T$ phase diagram.

Fig. 2: Scaling plot for the helicity moduli shown in Fig. 1(a) for $w = 0$ with $T_b = 0.8458J/k_B$, $\nu_5 = 0.728$, $\phi = 1.387$, and $\nu_2 = 0.666$. Inset (a): scaling plot based on Eq. (3); inset (b): magnification of the main panel around zero helicity modulus.

Fig. 3: Scaling plot for the Néel order parameter shown in Fig. 1(b) for $w = 0$ with $T_b = 0.8458J/k_B$, $\beta_5 = 0.400$, $\phi = 1.387$, and $\beta_3 = 0.363$. Inset (a): temperature dependence of 5-component superspin order parameter; inset (b): magnification of the main panel around zero Néel order parameter.

Fig. 4: Scaling plot for the helicity moduli shown in inset (a) for $w = 0.1J$ with $g_b = 0.011J$ and the same $T_b$ and critical exponents in Fig. 2. Inset (b): magnification of the main panel around zero helicity modulus.

Fig. 5: Scaling plot for the Néel order parameter shown in inset (a) for $w = 0.1J$ with $g_b = 0.011J$ and the same $T_b$ and critical exponents in Fig. 3. Inset (b): magnification of the main panel around zero Néel order parameter.

Fig. 6: $g - T$ phase diagram with a tetracritical point for $w = -0.1J$. Inset (a): temperature dependence of the helicity moduli; inset (b): that of the Néel order parameters.

Fig. 7: Scaling plot for the helicity moduli shown in Fig. 6(a) for $w = -0.1J$ with $T_t = 0.846J/k_B$, $\nu_t = 0.721$, $\phi_t = 1.363$, and $\nu_2 = 0.666$. Inset (a): scaling plot based on Eq. (3); inset (b): magnification of the main panel around zero helicity modulus.

Fig. 8: Scaling plot for the Néel order parameter shown in Fig. 6(b) for $w = -0.1J$ with $T_t = 0.846J/k_B$, $\beta_t = 0.399$, $\phi_t = 1.363$, and $\beta_3 = 0.363$. Inset (a): temperature dependence of 5-component superspin order parameter; inset (b): magnification of the main panel around zero Néel order parameter.

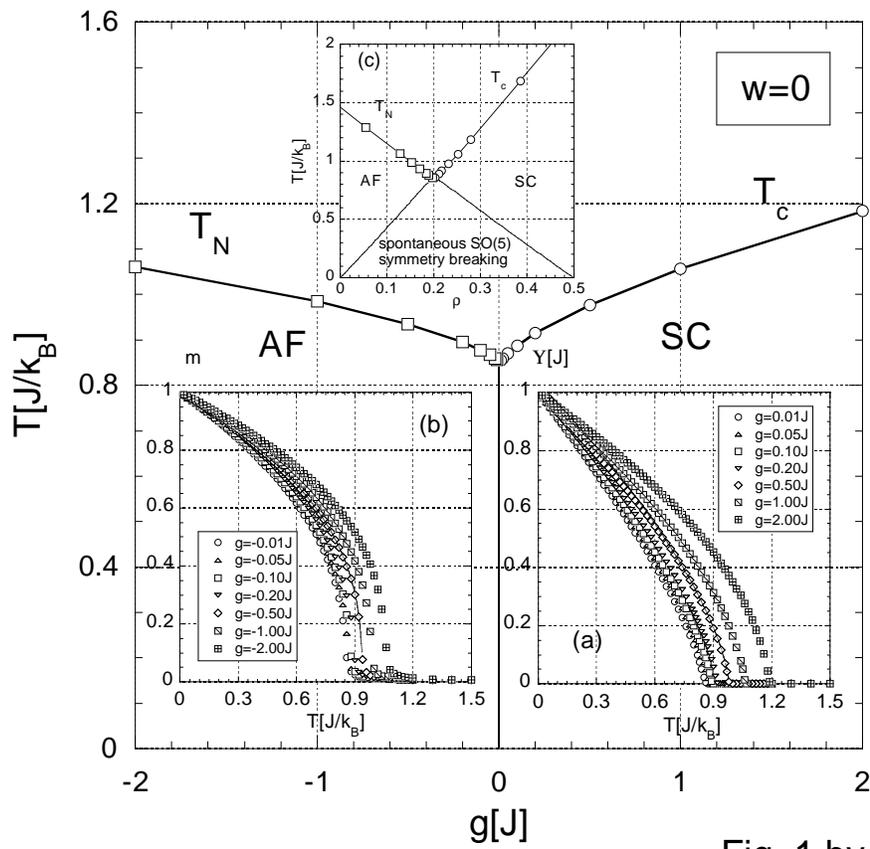

Fig. 1 by Xiao Hu

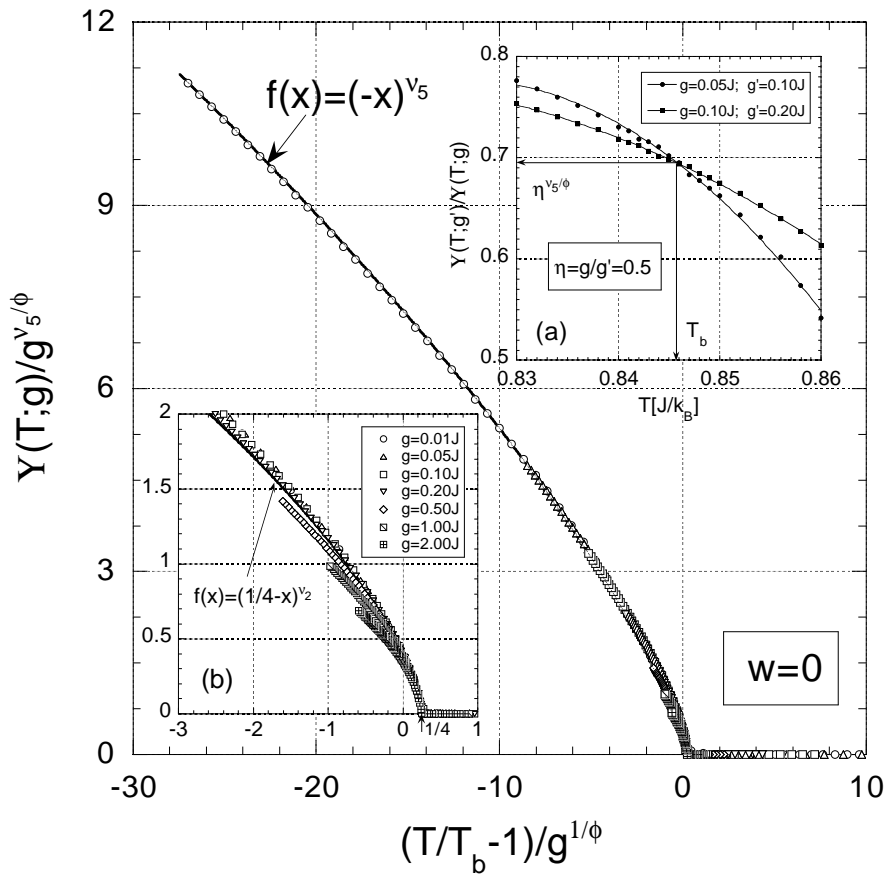

Fig. 2 by Xiao Hu

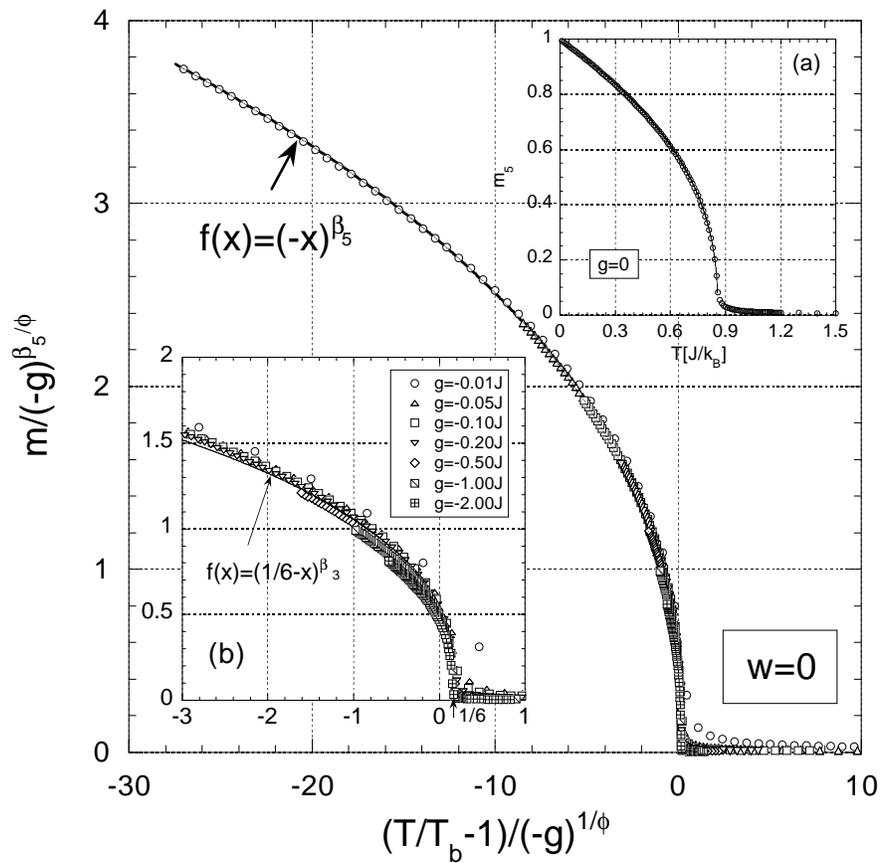

Fig. 3 by Xiao Hu

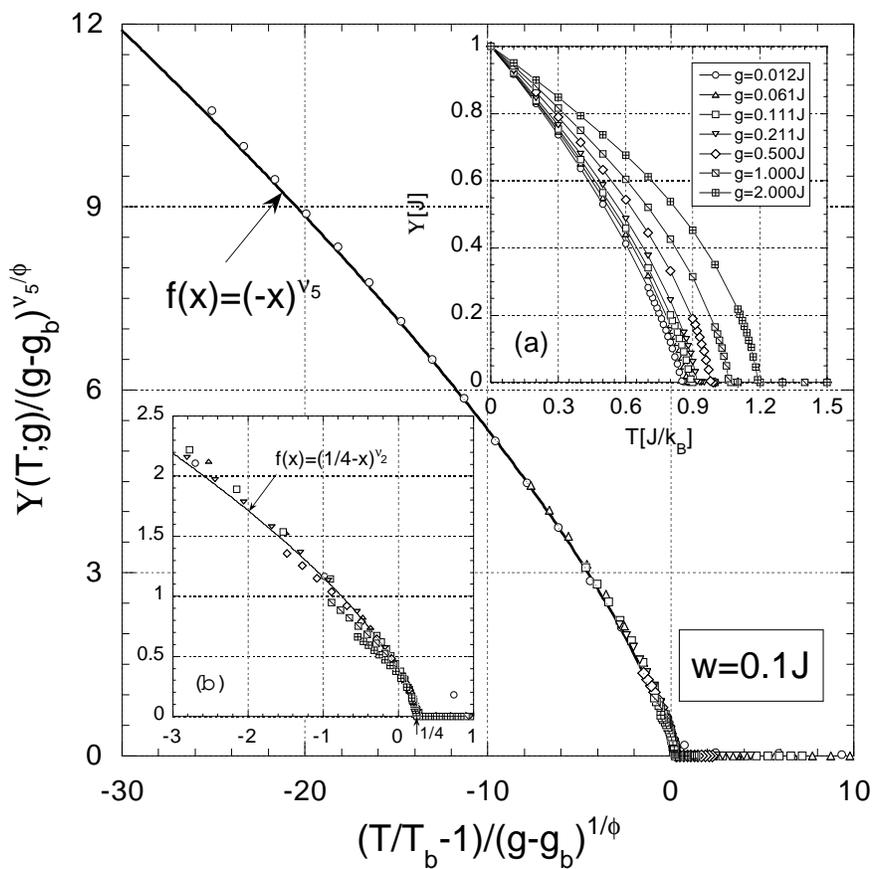

Fig. 4 by Xiao Hu

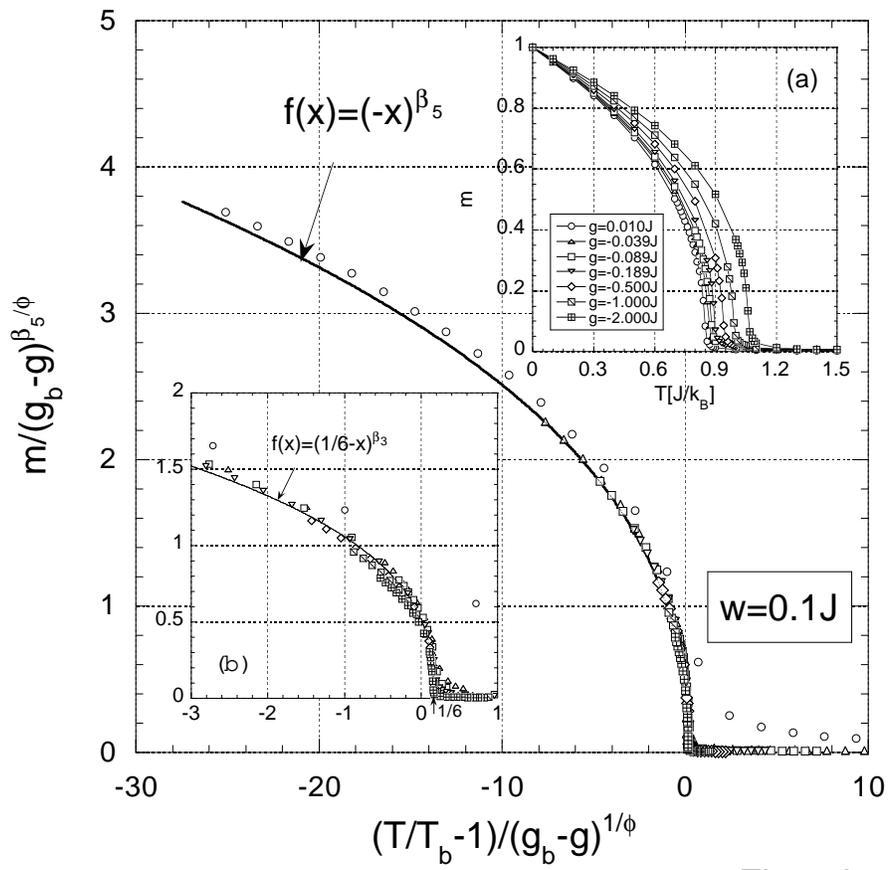

Fig. 5 by Xiao Hu

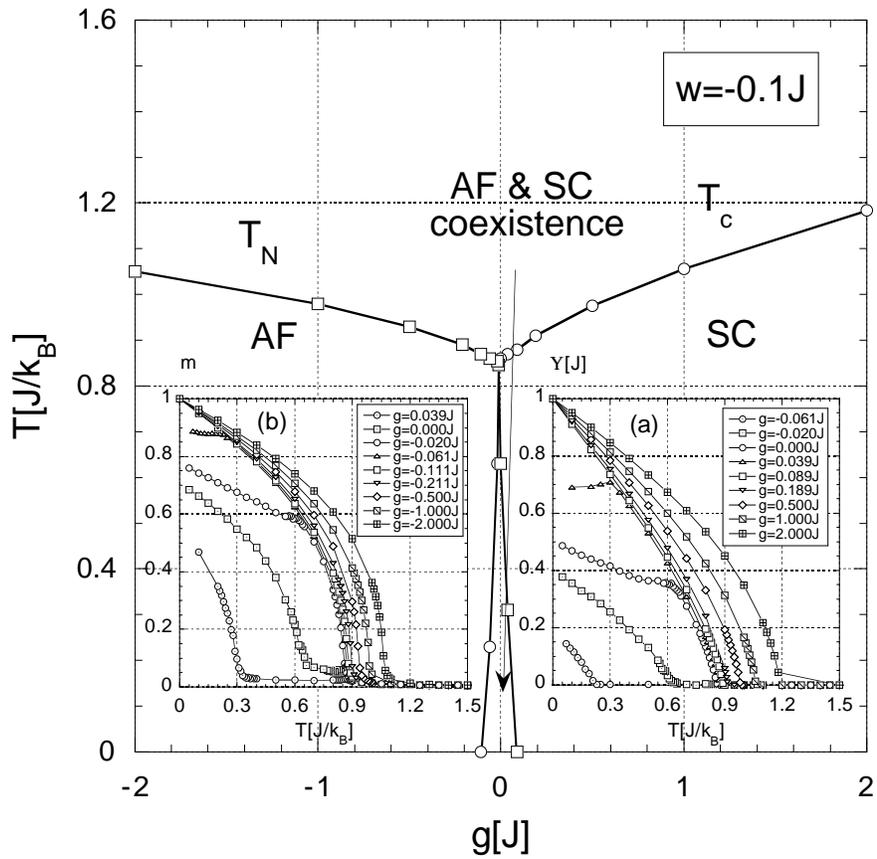

Fig. 6 by Xiao Hu

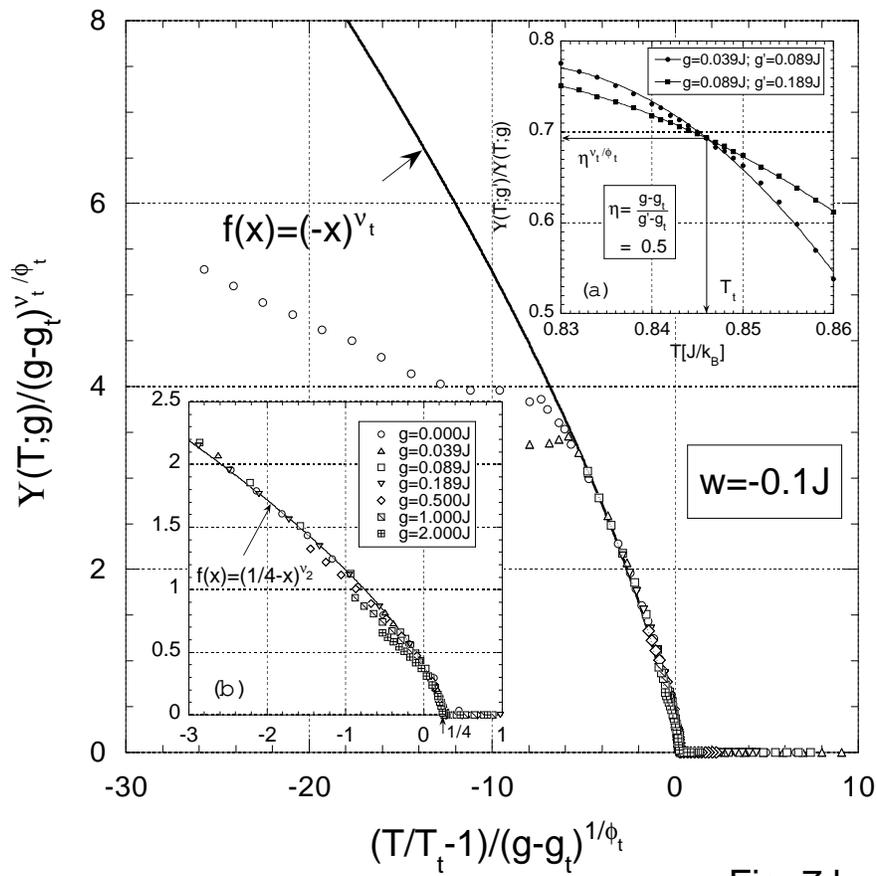

Fig. 7 by Xiao Hu

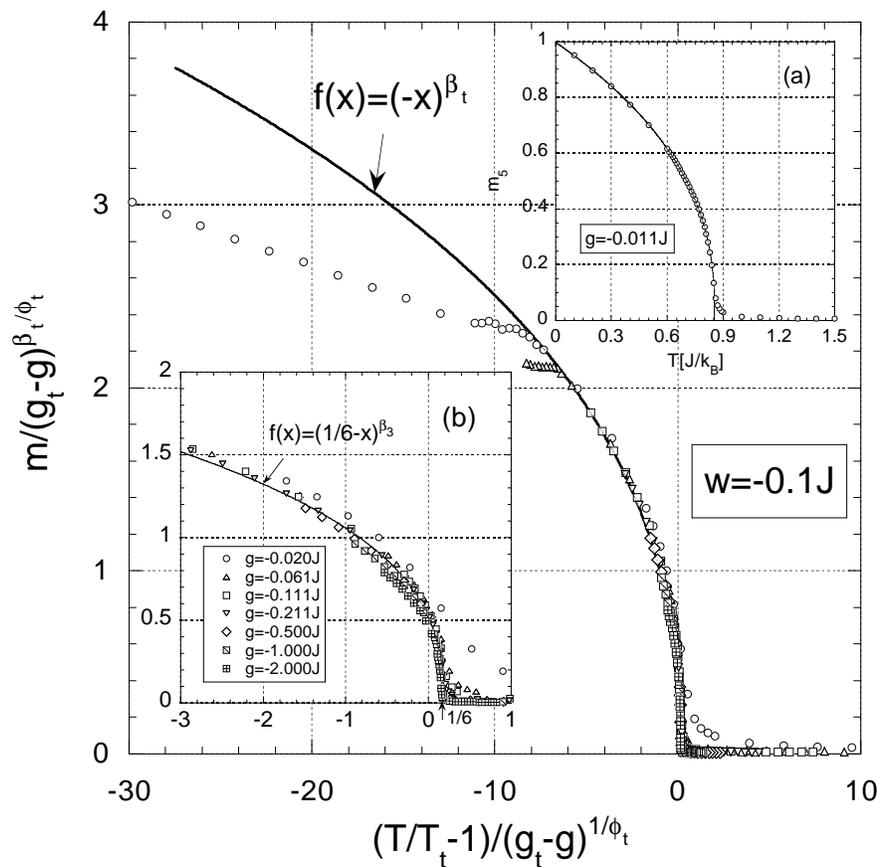

Fig. 8 by Xiao Hu